\documentclass[journal=jacsat,manuscript=article]{achemso}

\usepackage[version=3]{mhchem} 
\usepackage{bbold}

\author{Austin McDannald}
\affiliation[NIST MMSD]
{Materials Measurement Science Division, National Institute of Standards and Technology, Gaithersburg MD, USA}
\email{austin.mcdannald@nist.gov}
\alsoaffiliation{ORCID: 0000-0002-3767-926X}

\author{Daniel W. Siderius}
\affiliation[NIST CSD]
{Chemical Sciences Division, National Institute of Standards and Technology, Gaithersburg MD, USA}
\alsoaffiliation{ORCID: 0000-0002-6260-7727}

\author{Brian DeCost}
\affiliation[NIST MMSD]
{Materials Measurement Science Division, National Institute of Standards and Technology, Gaithersburg MD, USA}
\alsoaffiliation{ORCID: 0000-0002-3459-5888}

\author{Kamal Choudhary}
\affiliation[NIST MSD]
{Materials Science and Engineering Division, National Institute of Standards and Technology, Gaithersburg MD, USA}
\alsoaffiliation{ORCID: 0000-0001-9737-8074}

\author{Diana L. Ortiz-Montalvo}
\affiliation[NIST MMSD]
{Materials Measurement Science Division, National Institute of Standards and Technology, Gaithersburg MD, USA}
\alsoaffiliation{ORCID: 0000-0001-7293-4476}

\title[Intrinsic DAC]
  {Intrinsic Direct Air Capture}


\begin{document}







\begin{abstract}

We present new metrics to evaluate solid sorbent materials for Direct Air Capture (DAC). 
These new metrics provide a theoretical upper bound on \ce{CO2} captured per energy as well as a theoretical upper limit on the purity of the captured \ce{CO2}. 
These new metrics are based entirely on intrinsic material properties and are therefore agnostic to the design of the DAC system. 
These metrics apply to any adsorption-refresh cycle design.
In this work we demonstrate the use of these metrics with the example of temperature-pressure swing refresh cycles.
The main requirement for applying these metrics is to describe the equilibrium uptake (along with a few other materials properties) of each species in terms of the thermodynamic variables (e.g. temperature, pressure). 
We derive these metrics from thermodynamic energy balances. 
To apply these metrics on a set of examples, we first generated approximations of the necessary materials properties for 11\,660 metal-organic framework materials (MOFs). 
We find that the performance of the sorbents is highly dependent on the path through thermodynamic parameter space. 
These metrics allow for: 1) finding the optimum materials given a particular refresh cycle, and 2) finding the optimum refresh cycles given a particular sorbent. 
Applying these metrics to the database of MOFs lead to the following insights: 1) start cold – the equilibrium uptake of \ce{CO2} diverges from that of \ce{N2} at lower temperatures, and 2) selectivity of \ce{CO2} vs other gases at any one point in the cycle does not matter – what matters is the relative change in uptake along the cycle.

\end{abstract}

\section{Introduction}
There is already too much carbon in the atmosphere.\cite{IPCC-AR6}
Even if future \ce{CO2} emissions were avoided, there is still a need to remove the \ce{CO2} currently present to avoid the worst effects of climate change.\cite{IPCC-AR6}
One developing strategy is to use sorbents to directly capture the \ce{CO2} from the atmosphere for their subsequent sequestration and storage, known as direct air capture (DAC).

There are pilot DAC facilities, such as ClimeWorks, Carbon Engineering, and Global Thermostat, that have recently started operations.\cite{McQueen2021} 
However, there is still a need to drastically scale this technology to have the necessary impact. 
At the facility scale, the performance metrics and design considerations of DAC are well defined.
These include:
\begin{itemize}
    \item The \ce{CO2} Capture Efficiency: How much \ce{CO2} is captured per energy used.
    \item The \ce{CO2} Capture Output: How much \ce{CO2} is captured in a given length of time.
    \item The Refresh Cycle Time: The time it takes to perform one refresh cycle.
    \item The Purity of the Captured \ce{CO2}: The concentration of the \ce{CO2} in the output.
    \item The Sorbent Stability: The structural stability - especially (for DAC applications) in the presence of water vapor.
    \item The Sorbent Synthesizability: The ease and economic viablity of synthesizing the sorbent on an industrial scale. 
    \item The Sorbent Longevity: The time or number of refresh cycles the sorbent can experience before degrading to the point of needing to be replaced. 
\end{itemize}

However, many of these performance metrics are difficult to calculate or predict without detailed simulations of a particular process or the construction of a pilot plant. 
It can often be unclear what sorbent material properties will lead to good facility performance. 
In the synthesis of solid sorbents, much emphasis was put on achieving the synthesis of materials with high gravimetric or volumetric surface areas.\cite{Farha2012, Martin2013}
The motivation for this seems to be that higher surface areas should allow for more interactions, since the sorbent interacts with the sorbate (gas to be adsorbed) at the surface.
But these purely geometric measures incorporate no information about the interaction between the sorbent and sorbate. 
Another popular characteristic for comparing sorbents is working capacity. This is the difference in equilibrium uptake gas between two conditions - the amount of gas adsorbed and then desorbed on one refresh cycle. 
Working capacity does provide some information about the interaction between the sorbent and the sorbate, but working capacity is not enough to determine if a material will perform well as a sorbent. 
A sorbent with a small working capacity - but with little energy needed to cycle, or that could be cycled quickly - could be much better in a DAC facility than one with a large working capacity. 
Furthermore, working capacity is typically presented for conditions with a single sorbate (e.g. a pure \ce{CO2} environment at different partial pressures). 
Air, however, is a mixed gas. 
Therefore, the mixed gas adsorption behavior must be considered in DAC applications. 

There have been efforts to develop performance indicators for sorbents.
Jain \textit{et al.}\cite{Jain2003} developed a set of heuristics based on parameters such as particle size and time in the adsorption bed to aid in the design of pressure swing adsorption systems. 
Neumann \textit{et al.}\cite{Neumann2022} use a detailed model of the adsorption process in a gas separation column to calculate performance metrics for the sorbent. 
Similarly, Young \textit{et al.}\cite{Young2023} developed a machine learning model as a surrogate for detailed simulations of adsorption columns which allows them to screen many materials for that process, and provide insights into the sorbent and process design.   
While these can provide an accurate estimate of the performance metrics for that process, it depends on particular design choices of the adsorption process - such as the feed gas velocity, adsorption column design, and packing density. 
Ajenifuja \textit{et al}.,\cite{Ajenifuja2020}  created a model for screening sorbents for temperature swing based adsorption.
While this model is more directly relevant than using the selectivity or working capacity, it assumes that the adsorption system is a packed powder bed and, therefore, depends on extrinsic parameters such as the packing density. 
It further assumes that the refresh cycle is purely a temperature swing, which limits the generalizability to other thermodynamic spaces and can therefore not account for different paths through those spaces during refresh cycles, such as the increasingly popular temperature-vacuum swing adsorption cycles,\cite{Zhu2022} or to electro-swing adsorption cycles.\cite{ESAreview}. 
Recently, Charalambous \textit{et al}.\cite{holistic-platform} developed a holistic platform for evaluating carbon capture systems.
They do this by not only considering some of the aforementioned traditional sorbent performance indicators, but also additionally considering performance indicators of the process, a techno-economic assessment, and a life-cycle assessment. 
At the materials level, they primarily use the ratio of Henry's constants to indicate the performance, then feed the materials properties information into detailed simulations of an adsorption column process. 
The performance metrics we present in this work could readily be included into the platform developed by Charalambous \textit{et al}.\cite{holistic-platform} at the materials level to provide more informative sorbent performance indicators, and aid in the design of the processes - without the need to perform detailed process simulations. 
In general, the previous methods evaluating the performance of adsorption systems depend on extrinsic factors (such as the design of the gas separation column and consequential fluid dynamics). 
Furthermore, these previous methods typically only apply to a particular choice of refresh cycle - they would be difficult to generalize to other choices of refresh cycle, or even to other choices of process parameters. 

In this work, we develop sorbent performance metrics based on intrinsic material properties. 
These metrics directly generalize to any gas adsorption process when described in terms of thermodynamic parameters. 
Specifically, we develop the theoretical upper limit on the amount of \ce{CO2} captured per unit energy (which we term the Capture Efficiency) and the purity of the captured \ce{CO2}.
For illustrative purposes, we derive this model using the example of temperature vacuum swing adsorption cycle, which we visualize with an idealized piston. 
While this model is not how DAC facilities are likely to operate in practice, this framing of the problem elucidates all the relevant thermodynamic terms.
Much in the same way the Carnot cycle is a theoretical upper limit on the efficiencies of heat engines, these metrics are the theoretical upper limit of the capture efficiency and purity for gas separation processes based on adsorption.  
Inverting the capture efficiency is equivalent to estimating the theoretical lower limit of the energy cost needed to capture an amount of \ce{CO2}. 
For the purposes of this work, we will approximate air as 400 $\mu$mol/mol of \ce{CO2} with the balance of \ce{N2}.
While this will not capture the true behavior of the materials in air, especially the effects of \ce{H2O}, the analysis shown here on the binary approximation could be easily extended to additional components if provided a source of adsorption data. 
The rest of this document is organized into two parts: generating the necessary materials data needed for the analysis, and the thermodynamic analysis of an intrinsic DAC cycle.

\section{Generating the Data}
The goal of the metrics we develop in these works is provide a fair comparison between sorbent materials, and between thermodynamic refresh cycles. 
As such the thermodynamic analysis we use is based on intrinsic materials properties.  
Chief among those is the equilibrium uptake of each of the gas species in the mixture.
The heat of adsorption for each gas species is needed to keep track of the heat requirements. 
Because many refresh cycles include a temperature swing, in order to determine the energy requirements to change the temperature, the heat capacities ($C_V$) of sorbent materials are also needed. 
As a set of example materials, we generate all the needed materials property data for the 11\,660 metal-organic framework materials (MOFs) in the Cambridge Structural Database (CSD) database.\cite{CSD}
It is worth noting that while we apply these metrics to MOFs, they could just as easily be applied to zeolites, covalent-organic framework materials, hybrid systems, or any other solid sorbent system. 

\subsection{Equilibrium Uptake} \label{EquilibriumUptake}
The equilibrium uptake of \ce{CO2} in a sorbent is a function not only of temperature ($T$) and the partial pressure of \ce{CO2} ($P_{CO_2}$), but also the partial pressure of the other gasses - in this case the partial pressure of \ce{N2} ($P_{N_2}$). 
That is $n_{CO_2}(T, P_{CO_2}, P_{N_2})$, and similarly $n_{N_2}(T, P_{CO_2}, P_{N_2})$. 

However, there is very little mixed gas adsorption data, \cite{opening_toolbox} and even single component adsorption measurements for most sorbents are not repeatable. \cite{repoducible_sorbents}
We therefore turn to simulations to find the equilibrium uptake. 
Since we are screening 11 600 materials, we approximate the equilibrium uptake by first calculating single component Henry's constant isotherms with Grand Canonical Monte Carlo (GCMC) simulations, extrapolating those to the relevant temperatures, and then using Ideal Adsorbed Solution Theory (IAST) to obtain the mixed gas equilibrium uptake. 

Since the typical pressures for DAC are so low (near atmospheric or lower), we can approximate the isotherms for many sorbents with a simple Henry's law constant ($K_H$).
We can calculate $K_H$ by performing the GCMC simulation at $P = 0$. 
We use Free Energy and Advanced Sampling Simulation Toolkit (FEASST) to implement all of our GCMC calculations,\cite{FEASSTinitial} which was recently updated.\cite{FEASSTupdate}
The $K_H$s can be found by always rejecting the Monte Carlo move and simply recording the interaction energies.  
As an approximation to atmospheric conditions, we will consider an input total pressure of 101.3 kPa (1 atm) with a \ce{CO2} concentration of 400 $\mu$mol/mol (400 ppm) with the balance of \ce{N2}.
For each sorbent considered, we therefore perform 2 GCMC calculations to find both $K_{H,CO_2}$, and $K_{H,N_2}$.
Under this approximation of atmospheric conditions the partial pressure of \ce{CO2} is likely well within the linear regime of the isotherm, whereas most of the total pressure is due to the balance of \ce{N2}.
Even in the case of a pure \ce{N2} atmosphere, we expect that for many sorbents even 101.3 kPa is in the linear regime of the isotherm, and can therefore be approximated well by $K_{H,N_2}$.
In order to verify that $K_{H,N_2}$ is a good approximation, after we calculate the $K_{H,N_2}$ of each sorbent, we then perform a another GCMC calculation to directly predict the uptake at $P_{N_2} = 101.3$ kPa, and compare the predicted uptakes. 
The $K_{H,N_2}$ was considered a good approximation when $\pm 10~\%$ of the uptake from that was within the $95~\%$ confidence interval of the direct GCMC calculation. 
Those materials where these two predictions do not agree will require more sophisticated approximations, and should be the subject of future studies. 

These $K_{H}$s can be extrapolated in inverse temperature ($\beta = \frac{1}{k_BT}$, where $k_B$ is Boltzmann's constant) to give the temperature dependence of the adsorption behavior.\cite{Siderius2022} 
The GCMC simulations for the $K_H$s were performed at sufficient statistics to confidently populate an order-20 polynomial in $\beta$. 
By construction of these GCMC simulations, the $K_H$ will monotonically decrease with increasing temperature. 
We therefore truncate the thermodynamic extrapolations to the temperature where this order-20 polynomial becomes non-monotonic. 
To determine a lower limit for this extrapolation we compare the extrapolated $K_H$ to the saturation adsorption.
Since \ce{CO2} will reach saturation at higher temperatures than \ce{N2}, we consider the adsorption saturation of \ce{CO2}.
To determine the adsorption saturation, we perform a fourth GCMC simulation at increasing fugacity until the equilibrium uptake of \ce{CO2} converges.
We truncate the extrapolation to temperatures above where $K_H * P_{CO_2}$ exceeds the saturation uptake, $n_{sat,CO_2}$.
This allows us to extrapolate the $K_H$ with uncertainty to arbitrary temperatures, covering the temperature swings we will consider for the refresh cycles. 

With the temperature dependence of the $K_H$s, we can now use IAST to obtain the mixed gas equilibrium uptake at arbitrary temperatures and partial pressures of each gas.
In this low pressure limit of linear $K_{H}$ isotherms, IAST can be solved analytically. 
In the supplemental material we show the derivation of the equilibrium uptake along the desorption path in Step 2 of the intrinsic refresh cycle.
We show this derivation for three cases: analytically solving IAST for the binary mixture of \ce{CO2} and \ce{N2}, a correction for finite volume of the desorption chamber, and the generic case of non-linear adsorption behavior in multi-component gas mixtures (given the known functions $n_A(T,P_A,P_B,P_C,...)$, $n_B(T,P_A,P_B,P_C,...)$, $n_C(T,P_A,P_B,P_C,...)$, where $A$, $B$, $C$, ... refer to arbitrary gas species).

This procedure allows us to determine both $n_{CO_2}(T, P_{CO_2}, P_{N_2})$ and $n_{N_2}(T, P_{CO_2}, P_{N_2})$.
Furthermore, the derivative of the order-20 polynomial of $K_H(\beta)$ provides heat of adsorption at infinite dilution ($q_{ads}^{\infty}$).

\subsection{Sorbent Heat Capacity} 

In Ref.~\citenum{Moosavi2022}, Moosavi \textit{et al.} trained machine learning models, specifically XGBoost\cite{XGBoost} models, to predict the $C_V$ of MOFs and zeolites. 
These models were trained against a computational workflow that involved density functional theory calculations and molecular dynamics simulations for the force of atom displacements, phonon modes of the material, and subsequent calculations of the $C_V$s.
The computational workflow allows for the integration of the phonon modes across temperature to obtain the temperature dependence of the $C_V$s. 
In their work, Moosavi \textit{et al.} used separate ensembles of XGBoost models to predict the $C_V$s at several different temperatures (250 K to 400 K in 25 K increments). 
In this work, we use the pre-trained ensembles of 100 XGBoost models for each temperature from Ref.~\citenum{Moosavi2022} and their code repository \cite{CvRepo} for the predictions and uncertainties of the $C_V$s.  

To interpolate and extrapolate the $C_V$ predictions to arbitrary temperatures, we use a heteroscedastic Gaussian Process regressor (hGPR), implemented using Ref. \citenum{fairbrother2022gaussianprocesses}. 
This hGPR propagates the uncertainties from the $C_V$ predictions from the XGBoost model to make new predictions of the $C_V$ with quantifiable uncertainty at arbitrary temperatures. 
This hGPR model therefore allows us to predict the $C_V$ of the MOF sorbents at each step of the refresh path discussed in the next section.

\section{Thermodynamic Model}

\subsection{Model}
\begin{figure}

\includegraphics[width = \textwidth]{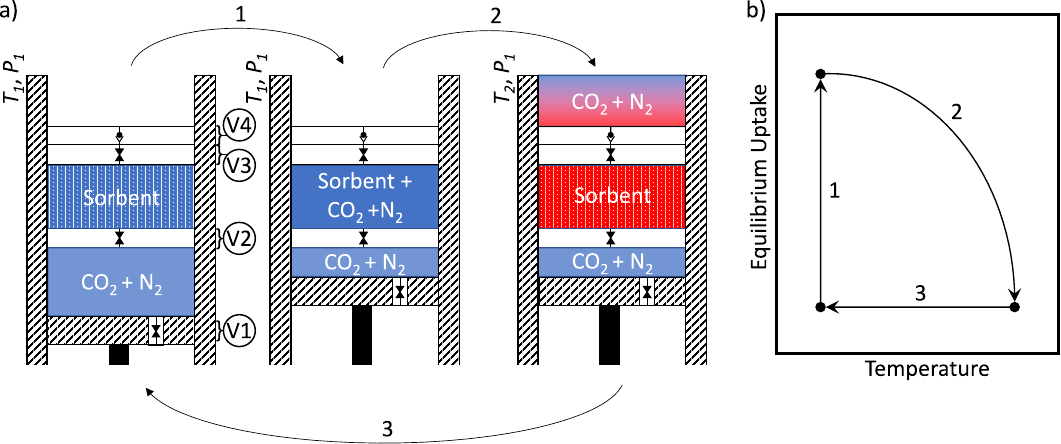}
\caption{a) A diagram of a piston model for adsorption process. 
This diagram shows an example temperature swing adsorption cycle. 
The diagonal hashed rectangles represent perfectly insulative walls. 
The valves are labeled V1 to V4. The steps 1 to 3 are steps of the example temperature swing adsorption cycle, starting at temperature $T_1$, ending at temperature $T_2$, at pressure $P_1$. 
The blue-to-red color scheme represents the low-to-high temperature. 
b) A sketch of the example temperature swing refresh path in temperature and equilibrium uptake space. 
The steps 1 to 3 correspond to those in part a). }
\label{fig:piston}
\end{figure}

In order to envision all the thermodynamic terms, it is helpful to consider a model. 
For this, we will use an idealized piston in Figure \ref{fig:piston}. 

This idealized piston system allows for the visualization of all of the terms in the energy and mass balances. 
Real DAC systems are typically packed columns of sorbent, and the performance of these systems depends on many extrinsic factors such as the column diameter, the flow rate, and the packing density in the column. 
By considering this idealized piston system, we avoid those factors and instead depend on intrinsic material properties. 
This idealized piston makes three strong assumptions: 1. the system reaches thermodynamic equilibrium at each step in the process, 2. the system does not consider any extrinsic factors like sorbent particle packing fraction, 3. as the gas is desorbed it leaves the system through a zero-volume check-valve as discussed in more detail below (see supplemental material for correction terms for this). 
With those assumptions, this system will find the theoretical upper limit on the capture efficiency as well as the purity of the captured gas. 
The utility of this work then is twofold.
The intrinsic refresh analysis can first be used to find optimal materials for a given thermodynamic process.
Secondly, this intrinsic refresh analysis can be used to find optimal thermodynamic processes for a material. 
In the derivations in the next section we show how this work could be extended to consider real-world extrinsic design considerations such as packing fraction and waste heat recovery. 

The surfaces of the piston chamber are assumed to be perfectly insulative; the piston is assumed to be mass-less, and the valves and check-valves are assumed to have no volume. 
The refresh cycle proceeds as follows:

\newcounter{desccounter}
\newcommand{\descitem}[1]{\refstepcounter{desccounter}\item[Step \thedesccounter{}#1:]}
\begin{description} 
    \descitem{} Valve 1 is closed and Valve 2 is opened, exposing the gas to the sorbent. The sorbent isothermally reaches the equilibrium uptake.
    \descitem{} Valve 2 is closed, Valve 3 is opened. The temperature and pressure are changed along a specified path through this thermodynamic space. The desorbed gas leaves through the outlet check-valve, Valve 4.
    \descitem{} Valve 3 is closed, Valve 1 is opened, and the piston is drawn back. The system returns to the initial temperature and pressure. 
\end{description}

To approximate atmospheric gas, we consider a binary mixture of \ce{N2} and \ce{CO2}. 
It is important to clarify that, for the purpose of all the subsequent analyses in this work, the model will use absolute adsorption as the thermodynamically relevant measure. 
Absolute adsorption includes not only the molecules of gas interacting with the surface of the sorbent, but also the molecules of gas in the pores of the sorbent. 
For a more in-depth discussion of the measures of adsorption, see Ref.\citenum{Brandani2016}.

\subsection{Energy Balance}
During Step 1, the (refreshed) sorbent is exposed to the new gas. 
For DAC this would be exposing it to the atmosphere, at ambient or slightly above ambient pressure. 
The sorbent adsorbs gas until it reaches equilibrium with the gas in the inlet source. 
Since adsorbing gas is exothermic, energy will be spent cooling the system to maintain the isothermal condition during adsorption. 
The gas being adsorbed changes the volume of the system.
There is a work term associated with this gas contracting.
Therefore the energy balance for Step 1 (assuming ideal gas behavior) is:
\begin{equation}
\label{eqn:e_balance_1}
\begin{aligned}
    E_1 &= Q_{ads,CO_2} + Q_{ads,N_2} - W_1 \\
    E_1 &= \Delta n_{CO_2}q_{ads,CO_2}(T_1) 
         + \Delta n_{N_2}q_{ads,N_2}(T_1) 
         + \Delta n_{CO_2} - P_1 \Delta V \\
    E_1 &= \Delta n_{CO_2}(q_{ads,CO_2}(T_1) + RT_1)
          +\Delta n_{N_2}(q_{ads,N_2}(T_1) + RT_1)
\end{aligned}
\end{equation}
Where $E_1$ is the energy balance of Step 1, $Q_{ads}$ is the heat of adsorption, $W_1$ is the work of the gas contracting, $P_1$ is the inlet pressure, $\Delta V$ is the change in volume of the gas, $\Delta n$ is the change in equilibrium uptake of the gas (i.e. the working capacity of that gas), $q_{ads}$ is the molar heat of adsorption, $T_1$ is the initial temperature, and $R$ is the ideal gas constant. 
Note that we know the initial conditions $(T_1,P_{1,CO_2},P_{1,N_2})$ and can measure or predict $q_{ads}(T)$ (as seen in the previous section), but do not yet know the $\Delta n$ terms. 
We can get the initial equilibrium uptake ($n_{1,CO_2}$ and $n_{1,N_2}$) from the initial conditions, but the equilibrium uptakes at the end of the cycle ($n_{end, CO_2}$, and $n_{end, N_2}$) depend on the path taken during the desorption in Step 2. 
This is because, while the total pressure $P_{end}$ can be chosen,
the composition of the desorbing gas changes throughout the refresh cycle since (under our modeling assumptions) it maintains equilibrium with the sorbent and remaining sorbate.
As we describe in more detail below, the composition of the captured gas at the end of the refresh cycle is determined by the integral over these small changes and is, therefore, highly path-dependent. 

During Step 2, the system is now isolated from the atmosphere, and the temperature and pressure are changed, causing gas to desorb from the sorbent. 
This desorption is endothermic, requiring energy to be put into the system. 
There is a work term associated with the gas expanding.
Energy is required to heat the gas in the system and the sorbent. 
Finally, there is the energy required to change the pressure. 
The total energy balance for Step 2 is therefore:
\begin{equation}
\label{eqn:e_balance_2}
\begin{aligned}
    E_2 &= Q_{ads,CO_2} + Q_{ads,N_2} - W_{2, CO_2} - W_{2, N_2}  \\
    & \ \ \ + E_{heat, CO_2} + E_{heat, N_2} + E_{heat, sorb} + E_{\Delta P}
\end{aligned}
\end{equation}

\noindent If we define a path $s$ through $(T,P)$-space, then we can define the desorption or refresh cycle as:
\begin{equation}
\label{eqn:path}
\begin{aligned}
    T(s) \ | \ \forall s, \ 0 \leq \frac{dT}{ds} \in \mathbb{R} \\
    P(s) \ | \ \forall s, \ 0 \geq \frac{dP}{ds} \in \mathbb{R}
\end{aligned}
\end{equation}
Meaning that $T$ and $P$ are both functions of the path $s$, under the constraint that for all $s$ the change in $T$ with respect to $s$ is a non-negative Real number and the change in $P$ with respect to $s$ is a non-positive Real number. Then we can use Eq. \ref{eqn:path} to define any temperature-swing, pressure-swing (including vacuum swing), or combinations thereof refresh cycles for the desorption. 
Under this definition of the desorption path, Eq. \ref{eqn:e_balance_2} becomes:
\begin{equation}
\label{eqn:e_balance_2_full}
\begin{aligned}
    E_2 = \int\limits_{s_{1}}^{s_{end}} \Bigl( & q_{ads,CO_2}(s)n_{CO_2}(s)
        + q_{ads,N_2}(s)n_{N_2}(s) \Bigr. \\
        \Bigl. &- RT(s) n_{CO_2}(s)
         -RT(s) n_{N_2}(s) \Bigr.\\
        \Bigl. &+ \frac{9}{2}RT(s)n_{CO_2}(s)
        + \frac{7}{2}RT(s)n_{N_2}(s)
        + C_{V,sorb}(s) T(s) \Bigr.\\
        \Bigl. &+ \bigl(n_{N_2}(s)+n_{CO_2}(s)\bigr) RT(s) log \bigl( P(s) \bigr) \Bigr) ds
\end{aligned}
\end{equation}
where $C_{V,sorb}(s)$ is the molar heat capacity at constant volume of the sorbent as a function of $s$. Note that here we assume that there is no change in volume of the sorbent with  $T$, $P$, or as gas is adsorbed/desorbed. 
We approximate the heat capacity of the adsorbed \ce{CO2} and \ce{N2} with the ideal gas theory approximation for triatomic and diatomic gases, respectively.

During Step 3, the system returns from $T_{end}$, and $P_{end}$ to $T_1$ and $P_1$. Since the aim of this work is to develop performance metrics that depend on the intrinsic properties of materials in order to directly compare potential sorbent materials, we assume no energy recovery. Therefore the energy balance for Step 3 is:
\begin{equation}
\label{eqn:e_balance_3}
    E_3 = 0
\end{equation}
However, real DAC systems will be able to recover energy from the hot output gas cooling and the total pressure equilibrating. A simple example of this is using heat exchangers for the hot gas on the output of one gas separation column to warm another column. Since those terms are system design dependent, we will ignore them for this analysis. 

The total energy balance for the refresh cycle is then simply the sum of the energies of each step:
\begin{equation}
\label{eqn:e_balance_total}
    E_{Total} = E_1 + E_2 + E_3
\end{equation}
We can then define our performance metrics for this intrinsic refresh cycle using the terms defined above. The purity of \ce{CO2} in the output is the mole fraction:
\begin{equation}
\label{eqn:purity}
    x_{end,CO_2} = \frac{\Delta n_{CO_2}}{\Delta n_{CO_2} + \Delta n_{N_2}}
\end{equation}
The intrinsic capture efficiency is how much \ce{CO2} was captured per unit energy: 
\begin{equation}
\label{eqn:capture_efficiency}
    \xi = \frac{\Delta n_{CO_2}}{E_{Total}}
\end{equation}
Alternatively, the inverse of $\xi$ is the energy cost to capture an amount of \ce{CO2}:
\begin{equation}
\label{eqn:cost}
    \mathcal{C} = \frac{1}{\xi} = \frac{E_{Total}}{\Delta n_{CO_2}}
\end{equation}

Note that in order to calculate $x_{end,CO_2}$ and $\xi$ we must first specify $s$, then determine $n(s)$ and $q_{ads}(s)$ for each gas species, as well as $C_{V,sorb}(s)$, which were shown in the previous section.

\section{Results and Discussion}
For ideal gases the enthalpy of mixing is zero - meaning that it takes zero work to separate ideal gases. 
This would mean that it would take zero work to capture the \ce{CO2} from the atmosphere. 
This would therefore imply that there is no upper limit on how much \ce{CO2} could be captured per energy.
However, the Gibb's free energy of mixing for ideal gases is negative, which means that the mixed gasses in atmosphere will not spontaneously separate. 
What we are modeling with the idealized piston described above is a way to keep track of all the thermodynamic terms for an adsorption based process.
Indeed, if we recovered all the waste heat in Step 3 - if we replace Eq. \ref{eqn:e_balance_3} to perfectly recover all the waste heat - then the total in Eq. \ref{eqn:e_balance_total} would necessarily sum to zero.
Real-world systems would likely employ some waste heat recovery and therefore have non-zero (but not perfectly efficient) $E_3$ terms, and could therefore exceed our stated upper limit of capture efficiency.
Since real-world waste heat recovery  systems are not perfectly efficient and would depend on the process design rather than materials properties, for the goal of comparing materials based on their intrinsic properties we ignore any heat recovery.
Therefore, this method provides an idealized upper bound on the purity and capture efficiency of sorbent based gas separations, in the absence of heat recovery - all of which is based off of intrinsic material properties. 

The most critical information for this analysis is the equilibrium uptake of each component gas as a function of temperature and the other component gases. 
Unfortunately there is very little data on adsorption behavior in mixed gases.\cite{opening_toolbox}
For the purposes of this initial work, we used thermodynamic extrapolation of $K_{H}$ isotherms and IAST to obtain the equilibrium uptakes.
These simplifying assumptions allowed us to screen through 11\,660 MOF materials found in the CSD database. 
However, the analysis here could just as easily be applied to sorbents with strongly non-linear isotherms calculated from higher levels of theory or from experimental measurements. 
All that is needed is a way to describe the equilibrium uptake of each gas species as a function of temperature and each partial pressure. 
However, as mentioned earlier, there is a significant lack of experimental measurements of mixed gas adsorption,\cite{opening_toolbox, repoducible_sorbents} and a lack of non-linear adsorption models that account for temperature dependence (extrapolating in temperature from isotherms). 
This work further assumes the volume of the sorbent does not change with adsorption.
Many MOF sorbents being considered for DAC applications are flexible and/or have nearly step-wise isotherms. 
To accurately consider these materials, the work associated with the sorbent volume change would need to be accounted for in the energy balance and specific heat calculations in Eqs. \ref{eqn:e_balance_1} and \ref{eqn:e_balance_2_full}. 

That being said, linear, non-interacting isotherms are good approximations for most MOF sorbents. 
For this study we considered the MOFs from the CSD.\cite{CSD}
Out of the 11\,600 materials studied in this work, there were 8\,759 materials where the $K_{H}$ are good approximations of the isotherms.
This was determined by comparing the equilibrium uptake of \ce{N2} at 101.3 kPa (1 atm.) and 300 K using the $K_{H,N_2}$ versus a direct GCMC calculation. 
A table categorizing all the completion mechanisms of our analysis is given in the supplemental material. 

\begin{figure}
\includegraphics[width = \textwidth]{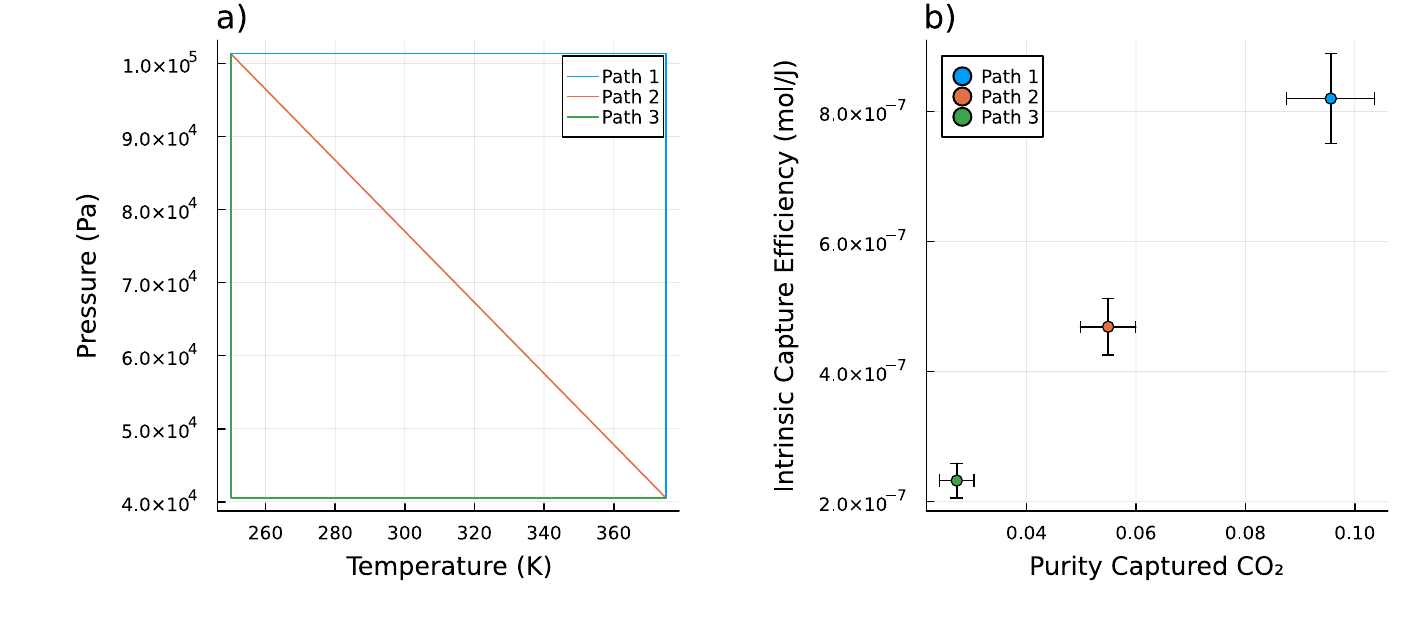}
\caption{a) A plot of 3 paths through $(T,P)$-space for the refresh cycle.  
b) The performance metrics of the sorbent (LETQAE01\_ion\_b) along each of the 3 paths.}
\label{fig:threepaths}
\end{figure}

The performance of the sorbent depends on the path through thermodynamic space during the refresh cycle. 
Figure \ref{fig:threepaths}a shows three example paths through $(T,P)$-space during Step 2. 
Each of these paths has the same starting and ending conditions, but take different paths through $(T,P)$-space to get there: heating first then pulling vacuum (Path 1, blue line), heating and pulling vacuum simultaneously (Path 2, orange line), and pulling vacuum first then heating (Path 3, green line). 
In Figure \ref{fig:threepaths}b, we show the purity of the captured \ce{CO2} and the intrinsic capture efficiency for one sorbent material, LETQAE01\_ion\_b, for each of these paths.
For this material, the performance in both metrics is optimized by heating first then pulling vacuum. 
The reason that the performance is path dependent is that, as the gas desorbs at one infinitesimal point along the path it is in equilibrium with the gas that desorbed at the previous infinitesimal point in the path. 

\begin{figure}
\includegraphics[width = \textwidth]{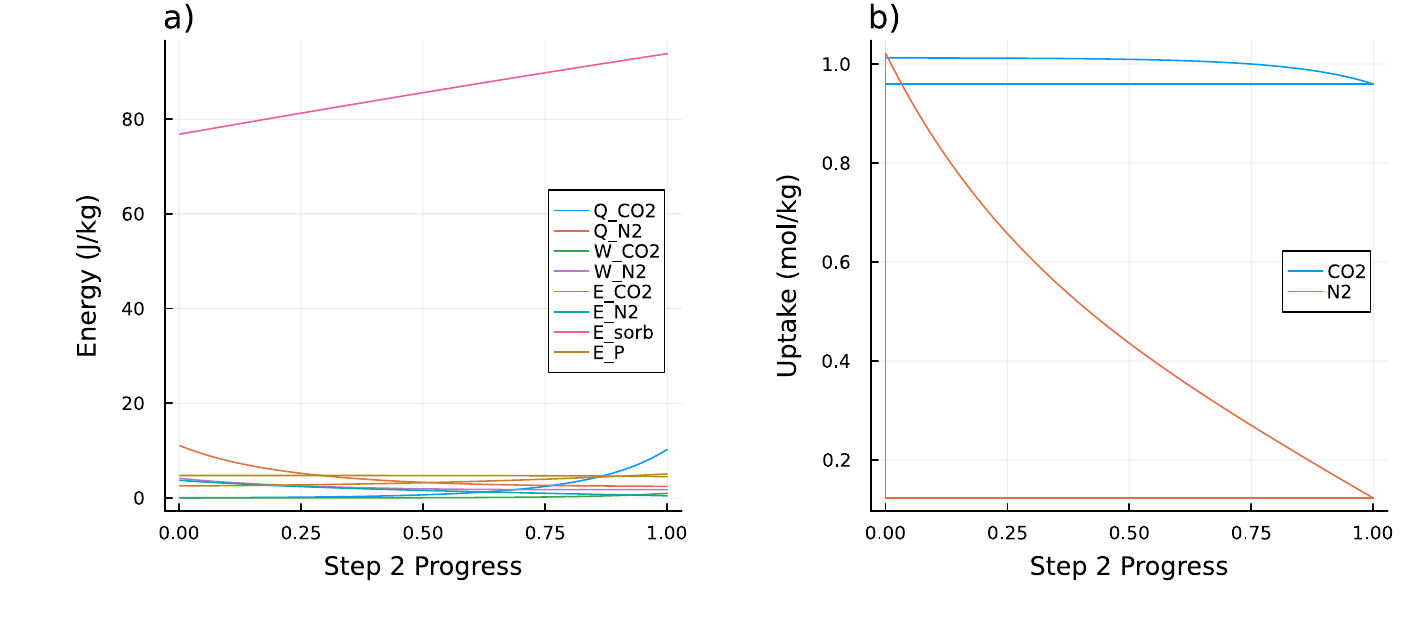}
\caption{A plot of each term in the energy balance (a), and the equilibrium uptake (b) for the sorbent LETQAE01\_ion\_b for Step 2 of the refresh cycle along path 2 from Figure \ref{fig:threepaths}a.  
Q\_CO2 and Q\_N2 are the heats of adsorption of \ce{CO2} and \ce{N2}, W\_CO2 and W\_N2 are the work associated with the change in volume of the desorbed \ce{CO2} and \ce{N2}, E\_CO2 and E\_N2 are the energies required to heat the adsorbed \ce{CO2} and \ce{N2}, E\_sorb is the energy required to heat the sorbent, and E\_P is the energy required to change the pressure.}
\label{fig:pathprogress}
\end{figure}

Figure \ref{fig:pathprogress}a shows each of the terms in the energy balance from Eq. \ref{eqn:e_balance_2_full} along the progress of Step 2 for Path 2 in Figure \ref{fig:threepaths}a. 
The heat of adsorption of \ce{N2} decreases as the cycle progresses, while the heat of adsorption of \ce{CO2} increases as the cycle progresses.
By far, the largest term in this energy balance is the energy required to heat the sorbent. 
Figure \ref{fig:pathprogress}b shows the equilibrium uptake of both \ce{CO2} and \ce{N2} along this refresh cycle, analogous to the diagram in Figure \ref{fig:piston}b. 
The equilibrium uptake along Step 1 is the vertical portion, while the equilibrium uptake along Step 3 is the horizontal portion. 
The working capacities are differences between equilibrium uptakes at the start and end of Step 2 (equivalently, the length of the vertical portion for Step 1). 

\begin{figure}
\includegraphics[width = \textwidth]{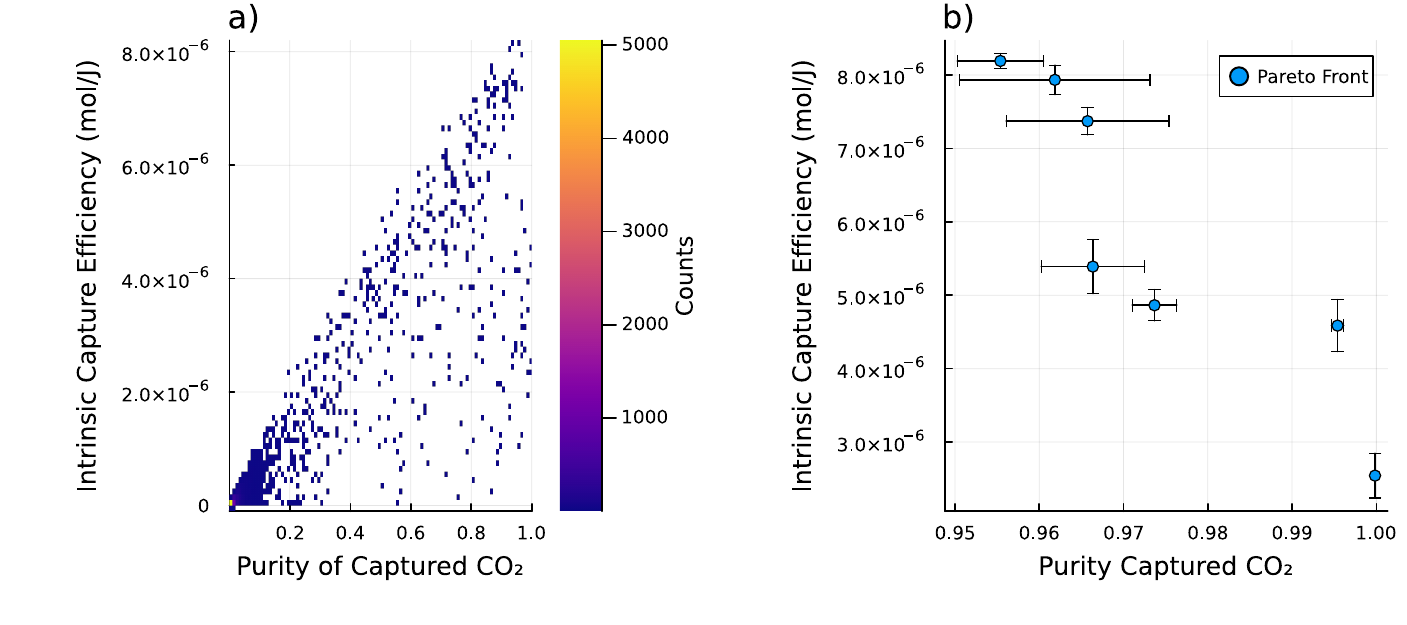}
\caption{a) A 2D histogram of the performance metrics of each MOF sorbent.  
b) A plot showing the Pareto front of all the sorbents given the specific path. }
\label{fig:SinglePath}
\end{figure}

We then consider a single path for the refresh cycle and examine the intrinsic refresh for all of the MOF materials in the CSD. This path is the same as Path 1 from Figure \ref{fig:threepaths}a - which starts at 250 K and 101325 Pa, warms isobarically to 350 K, then pulls vacuum isothermally to 40530 Pa. 
The 2D histogram of the performance metrics of all of the sorbents considering this path is shown in Figure \ref{fig:SinglePath}a. 
While there is a high density of sorbents that perform poorly by both metrics (toward the bottom left), there are still some sorbents that have impressive performance (toward the top right).
Figure \ref{fig:SinglePath}b shows the Pareto front of all the sorbents when considering this particular refresh path, which are also detailed in Table \ref{tab:ParetoOptimalSinglePath}.

\begin{table}
    \centering
    \resizebox{\linewidth}{!}{%
    \begin{tabular}{c|c|c|c|c}
    Name & $\xi$  & $x_{end}$ & $\Delta n_{CO_2}$ & $\Delta n_{N_2}$ \\
    \hline \hline
    jacs.6b06759\_ja6b06759\_si\_003\_clean & 8.191(99) & 0.9554(51) & 13.7(17) & 0.632(10) \\
    \hline
    ja5b02999\_si\_002\_clean & 7.93(20) & 0.962(11) & 12.6(32) & 0.4641(17) \\
    \hline
    RAVXIX\_clean & 7.37(18) & 0.9657(97) & 33.2(70) & 1.115(16) \\
    \hline
    PUPXII01\_clean & 5.39(37) & 0.9664(61) & 1.58(26) & 0.0534(13) \\
    \hline
    ZADDAJ\_clean & 4.86(21) & 0.9737(26) & 1.42(14) & 0.03802(92) \\
    \hline
    VAXHOR\_clean & 4.59(35) & 0.99539(74) & 1.05(17) & 0.00473(16) \\
    \hline
    BOMCUB\_charged & 2.54(30) & 0.999839(29) & 0.326(53) & 0.0000511(19) \\

    \end{tabular}}
    \caption{Table of the Pareto optimal materials given a single path (specifically Path 1 from Figure \ref{fig:threepaths}a). With intrinsic capture efficiency $\xi$ in ($\mu$mol/J), purity of the captured \ce{CO2} $x_{end}$, and working capacity $\Delta n$ in (mol/kg). Note that the number in the parenthesis represents the uncertainty in the previous two digits of the nominal value.}
    \label{tab:ParetoOptimalSinglePath}
\end{table}

\begin{figure}
\includegraphics[width = \textwidth]{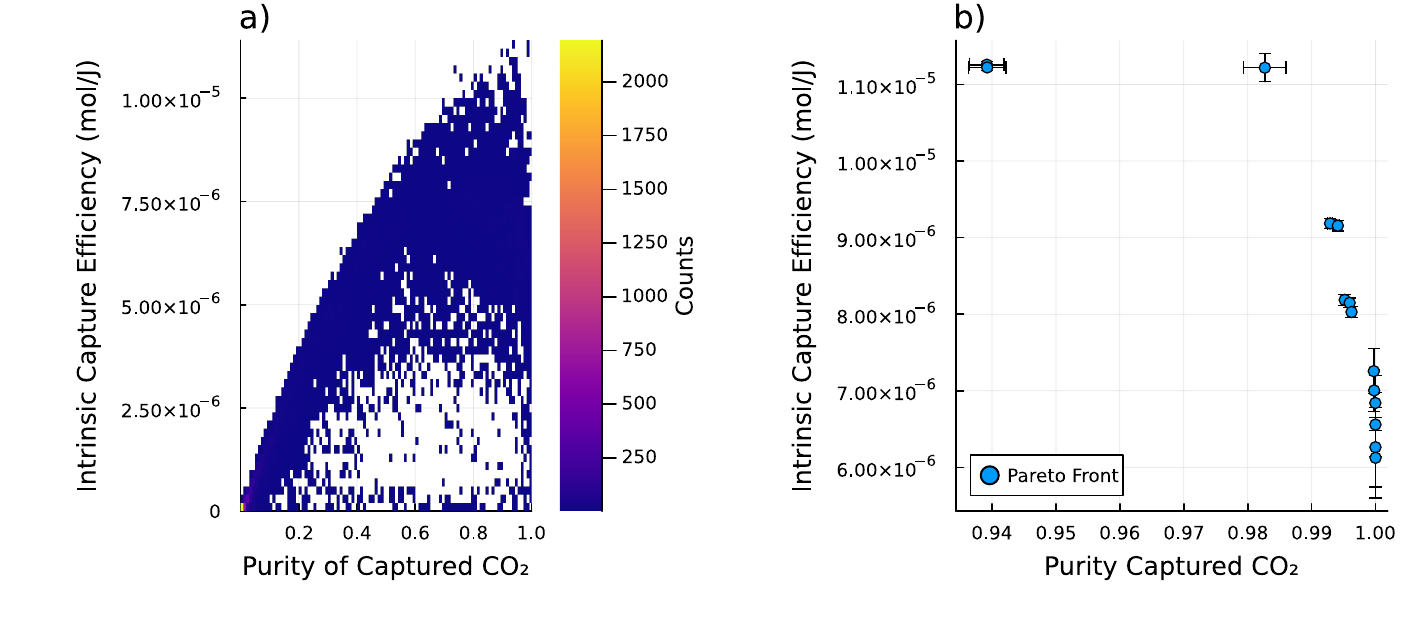}
\caption{a) A 2D histogram of the performance metrics of each MOF sorbent after finding the Pareto optimal paths.  
b) A plot showing the Pareto front of part (a). }
\label{fig:optimizedcycle}
\end{figure}

Next, we consider optimizing the path for each material.
The refresh path need only be monotonic in temperature and total pressure.
However, to simplify the parametrization of the path for the purposes of this study, we narrow the search space to only consider linear paths between starting point $(T_{1} , P_{1})$ and ending point $(T_{end} , P_{end})$. 
To enforce the monotonic paths we use the priors: 
\begin{equation}
\label{eqn:path_prior}
\begin{aligned}
    T_1 & \sim Uniform(200,400) \\
    T_{end} = T_1 + \Delta T \ | \ \Delta T &\sim Uniform(0, 200) \\
    P_1 & \sim Uniform(1.1*101325, 101325) \\
    P_{end} = P_1 + \Delta P \ | \ \Delta P &\sim Uniform(-101225,  0) 
\end{aligned}
\end{equation}
with $T$ in Kelvin, and $P$ in Pascal. 
We search for Pareto optimal refresh paths (Pareto optimal $T_1$, $\Delta T$, $P_1$, and $\Delta P$),  for each material in the database.
Figure \ref{fig:optimizedcycle}a shows a 2D histogram of the performance metrics of all the MOFs after finding the Pareto optimal paths.
The best performing materials are those in the top right of this figure. 
Despite the optimization there is high concentration of materials that do not perform well by either metric, which shows that these materials are likely poor candidates for DAC.
However, in comparison to Figure \ref{fig:SinglePath}a, there is a much higher density of sorbents that perform well.
This shows the utility of optimizing the refresh path for each sorbent. 
Figure \ref{fig:optimizedcycle}b shows the Pareto front of these materials and their corresponding paths, which are also summarized in Table \ref{tab:paretooptimal}. 
It is interesting to note that for many materials, refresh paths can be optimized to capture nearly pure \ce{CO2}. 
Furthermore, the capture efficiencies achieved after the path optimization are much higher than those seen in the single path considered in Figure \ref{fig:SinglePath}. 
There are a few sorbents with intrinsic capture efficiencies above $1.0 \times 10^5$ mol/J for their Pareto optimal paths.
To put some of the capture efficiency values in perspective, the average \ce{CO2} emissions by the US grid is about $2.46\times10^{-6}$ mol/J according to the US Energy Information Administration\cite{USGridEmmissions} - meaning at these idealized performances, many sorbents could potentially enable net-negative emissions with systems powered by the US grid. 
Of course these idealized efficiencies do not account for the real-world efficiencies of DAC systems - which would incur losses due to friction and other non-idealities, but could also include waste heat recovery. 
McQueen \textit{et al.}\cite{McQueen2021} states example systems currently operating with capture efficiencies of $2.27\times10^{-6}$ mol/J. 
However, that system was mostly powered by natural gas, which results in net positive emissions. 
In any case, this analysis provides an idealized upper limit on the capture efficiency which shows that real-world systems have potential for efficiency gains.

\begin{table}
    \centering
    \resizebox{\linewidth}{!}{%
    \begin{tabular}{c|c|c|c|c|c|c|c|c}
    Name & $\xi$  & $x_{end}$ & $\Delta n_{CO_2}$ & $\Delta n_{N_2}$ & $T_1$  & $T_{end}$  & $P_1$ & $P_{end}$  \\
    \hline \hline
     VANNIK\_clean   & 11.259(38) & 0.9392(27)  & 29.1(11) & 1.878(50) & 200.7 & 384.3 & 101950 & 60838 \\
     \hline
     VANNIK\_clean   & 11.223(41) & 0.9392(29) & 30.0(13) & 1.935(55) & 200.9 & 371.2 & 106080 & 29443 \\
     \hline
     LETQAE01\_ion\_b & 11.22(18) & 0.9827(33) & 137(24) & 2.346(30) & 200.2 & 391.2 & 101786 & 61381 \\
     \hline
     BEVQID\_clean   & 9.186(66) & 0.99287(37) & 3.17(13) & 0.02273(65) & 210.5 & 308.7 & 100868 & 10036\\
     \hline
     BEVQID\_clean   & 9.156(68) & 0.99409(29) & 3.75(16) & 0.02224(67) & 210.2 & 338.9 & 97326 & 7838 \\
     \hline
     FASJAL\_clean   & 8.186(69) & 0.99517(36) & 4.20(29) & 0.02032(58) &  236.6 & 316.7 & 99144 & 2432 \\
     \hline
     FASJAL\_clean   & 8.148(62) & 0.99594(23) & 4.64(25) & 0.01890(51) &  235.3 & 344.9 & 87503 & 4609\\
     \hline
     FASJAL\_clean   & 8.028(69) & 0.99623(22) & 4.78(24) & 0.01804(44) &  235.0 & 364.1 & 82466 & 3243\\
     \hline
     IFUDAO\_charged & 7.26(29) & 0.999715(46) & 1.61(22) & 0.000450(19) & 217.7 & 307.3 & 100360 & 9733\\
     \hline
     IFUDAO\_charged & 7.00(28) & 0.999767(31) & 2.02(24) & 0.000463(18) & 217.7 & 345.7 & 102382 & 14448\\
     \hline
     MAXHEA\_clean   & 6.84(36) & 0.9999511(83) & 0.99(15) & 0.0000471(26) & 213.1 & 300.3 & 110012 & 12457\\
     \hline
     MAXHOK\_clean   & 6.56(41) & 0.9999731(52) & 0.93(15) & 0.0000243(12) & 212.9 & 302.5 & 93225 & 9613\\
     \hline
     MAXHOK\_clean   & 6.26(52) & 0.9999739(63) & 1.04(20) & 0.0000259(12) & 213.8 & 331.7 & 101126 & 17833\\
     \hline
     PARFOF\_clean\_h & 6.12(52) & 0.9999911(19) & 1.14(21) & 0.00000974(45) & 203.6 & 311.7 & 102499 & 8995 \\

    \end{tabular}}
    \caption{Table of the Pareto optimal materials and their Pareto optimal paths. With intrinsic capture efficiency $\xi$ in ($\mu$mol/J), purity of the captured \ce{CO2} $x_{end}$, working capacity $\Delta n$ in (mol/kg), temperature $T$ in (K), and pressure $P$ in (Pa). Note that the number in the parenthesis represents the uncertainty in the previous two digits of the nominal value.}
    \label{tab:paretooptimal}
\end{table}

\begin{figure}
\includegraphics[width = 0.5\textwidth]{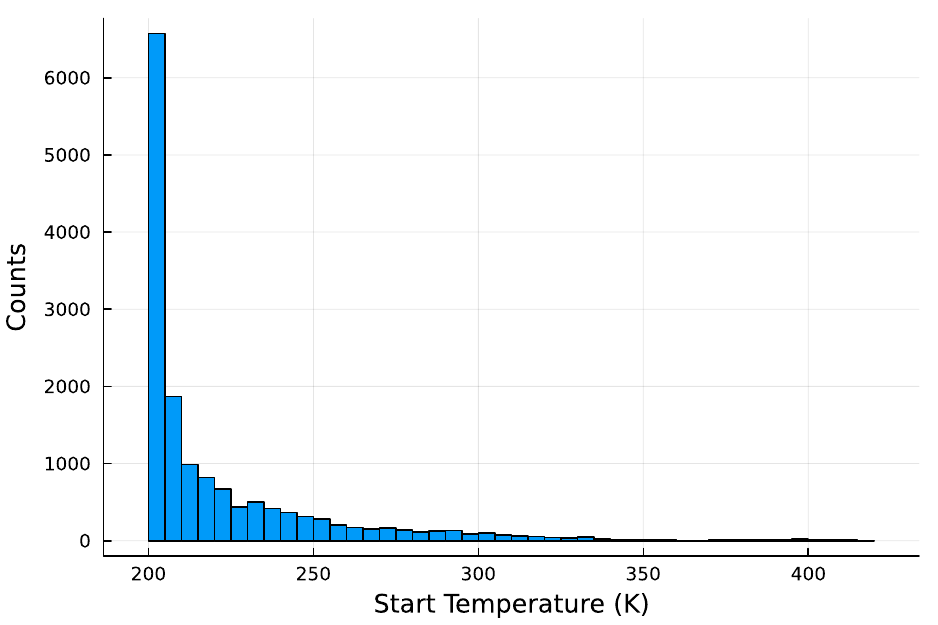}
\caption{A histogram of the starting temperature ($T_1$) of the Pareto optimal refresh paths.}
\label{fig:starttemphistogram}
\end{figure}

\begin{figure}
\includegraphics[width = \textwidth]{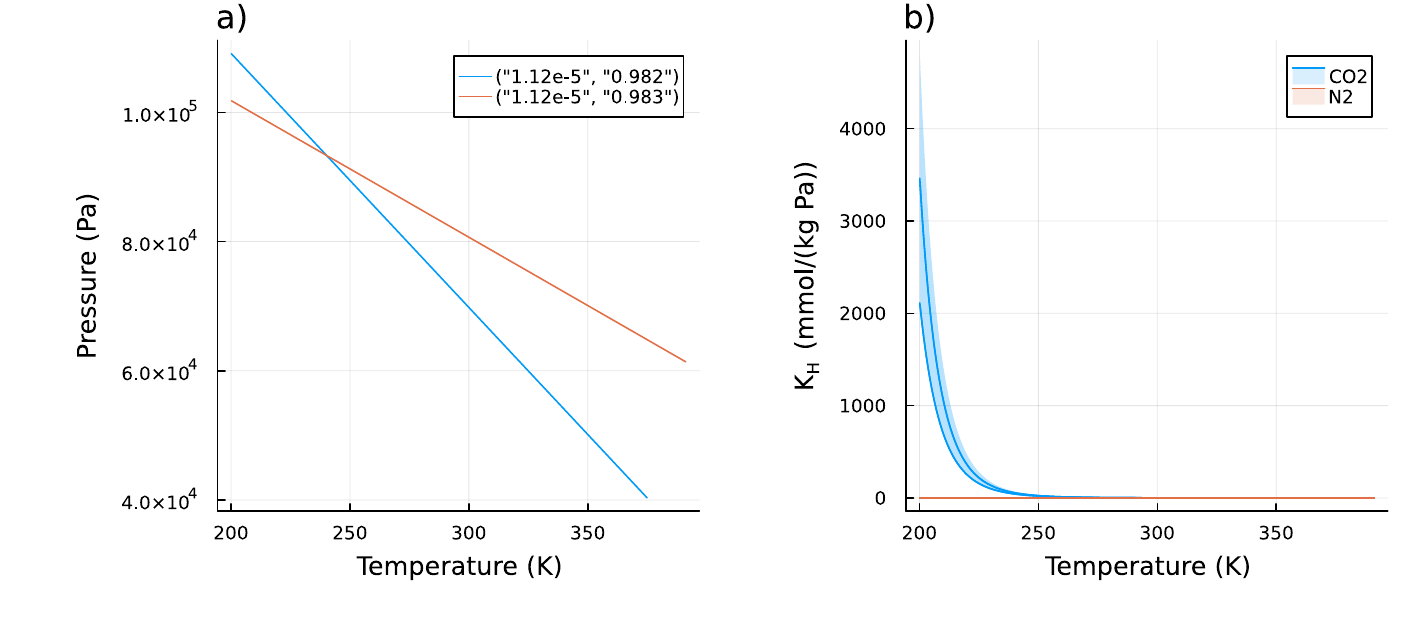}
\caption{a) The Pareto optimal refresh paths for LETQAE01\_ion\_b.
The legend labels the paths with the intrinsic capture efficiency (mol/J) and purity of captured \ce{CO2}, respectively.
b) The temperature dependence of the Henry's constant ($K_H$) for \ce{CO2} and \ce{N2}. 
The shaded region shows two standard deviations from the mean.}
\label{fig:inspectingpareto}
\end{figure}

One of the main conclusions from this analysis is that it is greatly beneficial to start cold.
Figure \ref{fig:starttemphistogram} shows a histogram of all the starting temperatures of all the Pareto optimal paths for all the materials considered. 
Most of the materials' adsorption performance were optimized by lowering the start temperature below ambient.
For this study we chose the lower limit of 200 K for the starting temperature to be somewhat outside of what many DAC systems would be designed to operate at, so as to not impose any undue restrictions. 
Yet the performance optimizations for many materials were hampered by this hard constraint.
We can gain some insight as to why the performance is optimized at low starting temperatures by inspecting the Pareto optimal paths of one material, LETQAE01\_ion\_b.
Figure \ref{fig:inspectingpareto}a shows the Pareto optimal paths for this material, while Figure \ref{fig:inspectingpareto}b, shows the temperature dependence of the Henry's constants for the relevant temperature ranges. 
There is a dramatic increase in the $K_{H,CO_2}$ as the temperature is lowered below 250 K, while the $K_{H,N_2}$ remains comparatively constant. 
From inspection of Eq. \ref{eqn:purity}, the purity is optimized when $n_{N_{2}}(s) = Constant $ or equivalently $ \Delta n_{N_{2}} = 0$. 
From inspection of Eqs. \ref{eqn:e_balance_1} and \ref{eqn:e_balance_2_full}, if $n_{N_{2}}(s) = Constant$ then the terms associated with the heat of adsorption of and work of adsorption of \ce{N2} tend toward zero, leaving only the small energy penalty associated with heating the adsorbed \ce{N2}. 
This means that if $n_{N_{2}}(s) = Constant$ then only \ce{CO2} is adsorbed or released, and some of the energy costs are eliminated. 
The implication of this is that selectivity for \ce{CO2} at any one point in the adsorption cycle is grossly incomplete information.
Nor is the working capacity of \ce{CO2} sufficiently informative of the performance. 
This is evident in Table \ref{tab:paretooptimal} where the working capacity of \ce{CO2} spans more than three orders of magnitude between the Pareto optimal materials despite fairly similar performance. 
The purity of the captured \ce{CO2} depends on the working capacity of \ce{CO2} and, crucially, the working capacity of \ce{N2}.
Since the dominant term in the energy balance is typically the heating of the sorbent material (as seen in Figure \ref{fig:pathprogress}), the capture efficiency is optimized by increasing the working capacity of \ce{CO2} and/or lowering the $C_{V,sorb}$.

\section{Conclusion}
In this work we developed performance metrics for sorbents for DAC based on intrinsic materials properties.
These metrics provide an idealized upper limit on the capture efficiency and the purity of the captured \ce{CO2}.
In order to evaluate these metrics the main information required is the equilibrium uptake of each of the gas species being considered as a function of the thermodynamic parameters (which include each independent partial pressure and temperature). 
In order to complete the energy balance, the sorbent materials properties that affect the relevant terms are needed. 
The most important of these - for refresh cycles involving temperature swings - is the sorbent heat capacity, which is typically the dominant contribution to the energy balance. 
In order to evaluate these metrics on the majority of the 11\,660 MOF materials found in the CSD database, we estimated the equilibrium uptake using IAST and GCMC calculated Henry's constants. 
We approximated atmospheric conditions with 400 $\mu$mol/mol \ce{CO2} with the balance of \ce{N2}.
We obtained estimates of the sorbent heat capacity at specific temperatures using pre-trained XGBoost from Ref.~\citenum{Moosavi2022}. 
We then interpolated and extrapolated these estimates to arbitrary temperatures using a heteroscedastic Gaussian Process regressor. 
These metrics can be used to find Pareto optimal sorbents for a given DAC refresh cycle - which could be useful in optimizing sorbent materials for existing DAC systems.
Alternatively, these metrics could be used to find the Pareto optimal DAC refresh cycle for particular sorbents - which could be used to design DAC systems to take advantage of available sorbents.
We used our simplified model to find the (linearly constrained) Pareto optimal refresh paths through $(T,P)$-space for the majority of the MOFs in the CSD database. 
With these results and an inspection of the energy balance it is clear that neither the selectivity at any one point along the cycle, nor the working capacity for \ce{CO2} are sufficient to describe the performance of the sorbent for DAC. 
We also show that due to the rapid change in the equilibrium uptake of \ce{CO2} at lower temperatures, it is beneficial to start the refresh cycles at lower temperatures. 
Many Pareto optimal refresh cycles start cold. 
We demonstrate the evaluation of these metrics for temperature-pressure swing refresh cycles, in a binary approximation of atmospheric conditions, and no waste heat recovery. 
However, it would be straightforward to include terms (in Eqs. \ref{eqn:e_balance_1} and \ref{eqn:e_balance_2_full}) for other refresh paths (e.g. electro-swing), for additional gas species (e.g. \ce{H2O}, \ce{O2}, as is shown in the supplemental material), or for energy recovery systems in real-world facilities (by adjusting Eq. \ref{eqn:e_balance_3}). 
Furthermore, while we used some simplified linear isotherms, many promising sorbents for DAC have highly non-linear (sometimes step-wise) adsorption curves.\cite{StepWiseAdsorption} 
As we show in the derivation presented in the supplemental material, these non-linear sorbents could be easily evaluated with the metrics we develop here \textit{if} the mixed-gas temperature-dependent adsorption behavior is well described, but unfortunately that data is not currently available for most materials.
This work, therefore, highlights the need for further studies on the equilibrium uptake as a function of temperature and in mixed gas environments.

\section{Disclaimer}
Any mention of commercial products in this report is for information only; it does not imply recommendation or endorsement by NIST.

\section{Author Contributions}
AM  initially conceptualized the study, developed the thermodynamic models and derived the metrics, developed the Julia code base for those metrics, ran GCMC and Intrinsic DAC cycle analysis calculations, and wrote the initial draft.
DWS wrote the FEASST scripts for the GCMC calculations, provided feedback on the equilibrium uptake and  thermodynamic models and derivations, and provided insight into the interpretation of the results.
BD helped develop the Julia code base.
KC and DLOM helped refine and guide the development of the study as part of the team led by DLOM. 
All authors contributed to the editing of the manuscript. 





\begin{suppinfo}

The supplemental information shows a derivation of the equilibrium uptake along the refresh path for linear isotherms and IAST, a correction for the real-world case of finite volume of the desorption chamber, and the generic case for non-linear adsorption curves.
The code used to implement the intrinsic DAC cycle analysis is available as a Julia package at: \url{https://github.com/usnistgov/IntrinsicDACCycle}
The results of the GCMC calculations and intrinsic DAC cycle calculations used in this study are available at \url{https://doi.org/10.5281/zenodo.14452152}. 

\end{suppinfo}

\bibliography{achemso-demo}

\end{document}




\section{Derivation Intrinsic DAC Cycle} \label{sec:derivation}
In this section we show the derivation of how we obtain the equilibrium uptake at each point along the desorption path (Step 2 from Figure 1a in the main text). 
In Section \ref{subsec:IAST} we show the derivation using Ideal Adsorbed Solution Theory.
In Section \ref{subsec:finiteV} we show a correction for the real-world case of a desorption chamber with finite volume.
Lastly, in Section \ref{subsec:genaric} we show the generalized derivation for generic functions for the equilibrium uptake and arbitrary number of species in the gas mixture.

\subsection{Derivation using Ideal Adsorbed Solution Theory} \label{subsec:IAST}
Initially, after the adsorption step, the material is in equilibrium with the inlet conditions. 
Then the material is isolated from the inlet and the thermodynamic conditions (i.e., temperature $T$, and total pressure $P$) are changed to desorb the gasses. 
In our idealized system, there is no volume to the outlet (an idealized check-valve) so the composition of gas that desorbs is the composition of the outlet (we show later, in Section \ref{subsec:finiteV} how to correct for this). 
Therefore, the equilibrium uptake and composition of gas that is desorbed by changing the conditions are determined from equilibrium with the material using Ideal Adsorbed Solution Theory (IAST). 
In the limit of low pressure, the equilibrium uptake of a gas is linear with partial pressure, the slope of which is the Henry Constant. 
As discussed in the main text, the conditions of Direct Air Capture refresh cycles are typically sufficiently low pressure that Henry Constant isotherms is a good prediction. 
The Henry constants are a function of temperature, and therefore change over the refresh cycle. 
IAST can be solved analytically with these linear, Henry Constant isotherms.

To determine the equilibrium uptake during desorption (and consequently the moles of each gas species in the outlet) along the refresh path through thermodynamic parameter space, we can iteratively solve the IAST equations for small step changes in the conditions.

\noindent \textbf{Given}: the initial Henry Constants for \ce{CO2} ($K_{H,1,CO_2}$), and \ce{N2} ($K_{H,1,N_2}$), initial Pressure $P_1$, and initial \ce{CO2} concentration $x_{1, CO_2}$. 

\noindent \textbf{Find}: the equilibrium uptake of \ce{CO2} and \ce{N2} during adsorption.
Then find the new \ce{CO2} concentration $x_{2, CO_2}$ during the desorption phase after isolating from the inlet and at the new conditions with $K_{H,2,CO_2}$, and $K_{H,2,N_2}$, and Pressure $P_2$.

\noindent The initial absolute equilibrium uptake of \ce{CO2} is:
\begin{equation}
\begin{aligned}
    n_{1,CO_2} &= K_{H,1,CO_2}P_{CO_2} \\
    &= K_{H,1,CO_2} P x_{1,CO_2} \\
\end{aligned}
\end{equation}

\noindent The initial absolute equilibrium uptake of \ce{N2} is:
\begin{equation}
\begin{aligned}
    n_{1, N_2} &= K_{H,1,N_2}P_{N_2} \\
    &= K_{H,1,N_2} P (1 - x_{1,CO_2})\\
\end{aligned}
\end{equation}

\noindent Next, the system is isolated from the inlet and the thermodynamic conditions are changed desorbing some of the gas. The change in equilibrium uptake of each species is:
\begin{equation}
\begin{aligned}
    d_{CO_2} &= n_{2, CO_2} - n_{1, CO_2} \\
    d_{N_2} &= n_{2, N_2} - n_{1, N_2} \\
\end{aligned}
\end{equation}
\noindent where $n_{2, CO_2}$ and $n_{2, N_2}$ are as yet unknown.

\noindent Since there is no volume of the outlet (idealized check valve), the change in equilibrium uptake (the gas that is desorbed: $d_{CO_2} + d_{N_2}$) is the gas of the outlet. The new concentration of \ce{CO2} is therefore:
\begin{equation} \label{eq:newconcentration}
\begin{aligned}
    x_{2, CO_2} &= \frac{d_{CO_2}}{d_{CO_2} + d_{N_2}}\\
    &= \frac{n_{2, CO_2} - n_{1, CO_2}}{ n_{2, CO_2} - n_{1, CO_2} + n_{2, N_2} - n_{1, N_2}}\\
    &= \frac{K_{H,2,CO_2}P_2x_{2, CO_2} - n_{1, CO_2}}{ K_{H,2,CO_2}P_2x_{2, CO_2} - n_{1, CO_2}  + K_{H,2,N_2}P_2(1-x_{2, CO_2}) - n_{1, N_2}}\\
\end{aligned}
\end{equation}

\noindent Rearranging Eq. \ref{eq:newconcentration}, we get:

\begin{equation}
\begin{aligned}
    K_{H,2,CO_2}P_2x_{2, CO_2} - n_{1, CO_2} = x_{2, CO_2}(& K_{H,2,CO_2}P_2x_{2, CO_2} \\
                                                          &- n_{1, CO_2} \\ 
                                                          &+ K_{H,2,N_2}P_2(1-x_{2, CO_2})\\ 
                                                          &- n_{1, N_2})
\end{aligned}
\end{equation}
\noindent Or:
\begin{equation} \label{eq:newconcentrationroot}
\begin{aligned} 
    0 = &(K_{H,2,CO_2}P_2 - K_{H,2,N_2}P_2)x_{2, CO_2}^2 \\
        &+ (K_{H,2,N_2}P_2 - K_{H,2,CO_2}P_2 - n_{1, CO_2} - n_{1, N_2})x_{2, CO_2} \\
        &+ n_{1, CO_2}
\end{aligned}
\end{equation}
This is the classic quadratic equation with:
\begin{equation}
\begin{aligned}
    A &= K_{H,2,CO_2}P_2 - K_{H,2,N_2}P_2\\
    B &= K_{H,2,N_2}P_2 - K_{H,2,CO_2}P_2 - n_{1, CO_2} - n_{1, N_2} \\ 
    C &= n_{1, CO_2}
\end{aligned}
\end{equation}
Therefore the new \ce{CO2} concentration is given by:
\begin{equation} \label{eq:root}
\begin{aligned}
    x_{2, CO_2} =\frac{-B \pm \sqrt{B^2 - 4AC}}{2A}
\end{aligned}
\end{equation}
Of the 2 roots to Equation \ref{eq:root} we choose the solution in the physically meaningful range: $x_{2, CO_2} \in [0,1]$ AND closest to $x_{1, CO_2}$ since each step is a small change in thermodynamic conditions.
We can then determine the new equilibrium uptake for \ce{CO2} and \ce{N2} with:
\begin{equation} \label{eq:end}
\begin{aligned}
    n_{2, CO_2} &= K_{H,2,CO_2}P_2x_{2, CO_2}\\
    n_{2, N_2} &= K_{H,2,N_2}P_2(1 - x_{2, CO_2})
\end{aligned}
\end{equation}

The equilibrium uptake during the desorption process can then be calculated for arbitrary paths through thermodynamic parameter space by iteratively solving Eq.s \ref{eq:root} and \ref{eq:end} for small steps. 
Note that this derivation assumes monotonic decreases in $P$ as well as monotonic decreases in $K_{H,CO_2}$ and $K_{H,N_2}$ along the refresh path.

\subsection{Derivation with Finite Volume} \label{subsec:finiteV}
Note that for real systems the outlet would have a known volume, which could be accounted for in Eq. \ref{eq:newconcentration} by including the appropriate terms in the nominator and denominator for the amount of \ce{CO2} in already in the outlet and total amount of gas already in the outlet, respectively. Here we introduce the new terms and show how to account for known finite volume of the outlet.

\noindent The initial absolute equilibrium uptake of \ce{CO2} is:
\begin{equation}
\begin{aligned}
    n_{1,CO_2} &= K_{H,1,CO_2}P_{CO_2} \\
    &= K_{H,1,CO_2} P x_{1,CO_2} \\
\end{aligned}
\end{equation}

\noindent The initial absolute equilibrium uptake of \ce{N2} is:
\begin{equation}
\begin{aligned}
    n_{1, N_2} &= K_{H,1,N_2}P_{N_2} \\
    &= K_{H,1,N_2} P (1 - x_{1,CO_2})\\
\end{aligned}
\end{equation}

\noindent The initial moles of \ce{CO2} in the free volume is:
\begin{equation}
\begin{aligned}
    m_{1,CO_2} &= \frac{P_{CO_2}V}{RT} \\
    &= x_{1,CO_2}\frac{PV}{RT} \\
\end{aligned}
\end{equation}

\noindent The initial moles of \ce{N2} in the free volume is:
\begin{equation}
\begin{aligned}
    m_{1,N_2} &= \frac{P_{N_2}V}{RT} \\
    &= (1-x_{1,CO_2})\frac{PV}{RT} \\
\end{aligned}
\end{equation}

\noindent The change in the equilibrium uptakes are:
\begin{equation}
\begin{aligned}
    d_{CO_2} &= n_{2, CO_2} - n_{1, CO_2} \\
    d_{N_2} &= n_{2, N_2} - n_{1, N_2} \\
\end{aligned}
\end{equation}

\noindent The new concentration considering the finite volume is:

\begin{equation} 
\begin{aligned}
    x_{2, CO_2} &= \frac{m_{1, CO_2} + d_{CO_2}}{m_{1, CO_2} + m_{1, N_2} + d_{CO_2} + d_{N_2}}\\
    &= \frac{m_{1, CO_2} + n_{2, CO_2} - n_{1, CO_2}}{ m_{1, CO_2} + m_{1, N_2} n_{2, CO_2} - n_{1, CO_2} + n_{2, N_2} - n_{1, N_2}}\\
\end{aligned}
\end{equation}

\noindent Substituting and re-arranging we get:

\begin{equation} 
\begin{aligned}
    0 = &(K_{H,2,CO_2}P_2 - K_{H,2,N_2}P_2)x_{2, CO_2}^2 \\
        &+ (\frac{P_1 V}{RT_1} + K_{H,2,N_2}P_2 - K_{H,2,CO_2}P_2 - n_{1, CO_2} - n_{1, N_2})x_{2, CO_2} \\
        &+ (n_{1, CO_2} - m_{1, CO_2})
\end{aligned}
\end{equation}


\noindent Similarly to Eq. \ref{eq:newconcentrationroot},  this is the classic quadratic equation with:
\begin{equation}
\begin{aligned}
    A &= K_{H,2,CO_2}P_2 - K_{H,2,N_2}P_2\\
    B &= \frac{P_1 V}{RT_1} + K_{H,2,N_2}P_2 - K_{H,2,CO_2}P_2 - n_{1, CO_2} - n_{1, N_2} \\ 
    C &= n_{1, CO_2} - m_{1, CO_2}
\end{aligned}
\end{equation}

\noindent With these definitions of $A$, $B$, and $C$, we can find the new \ce{CO2} concentration as before in Eq. \ref{eq:root} with:
\begin{equation} 
\begin{aligned}
    x_{2, CO_2} =\frac{-B \pm \sqrt{B^2 - 4AC}}{2A}
\end{aligned}
\end{equation}
As before, of the 2 roots we choose the solution in the physically meaningful range: $x_{2, CO_2} \in [0,1]$ AND closest to $x_{1, CO_2}$ since each step is a small change thermodynamic conditions.
We can then determine the new equilibrium uptake for \ce{CO2} and \ce{N2} with:
\begin{equation}
\begin{aligned}
    n_{2, CO_2} &= K_{H,2,CO_2}P_2x_{2, CO_2}\\
    n_{2, N_2} &= K_{H,2,N_2}P_2(1 - x_{2, CO_2})
\end{aligned}
\end{equation}

\subsection{Generic Intrinsic Refresh Cycle} \label{subsec:genaric}
In Sections \ref{subsec:IAST} and \ref{subsec:finiteV} we used linear Henry's constants and IAST to obtain the equilibrium uptake in the binary mixture of \ce{CO2} and \ce{N2}.
One advantage of using this framework, is that the uptake can be solved for analytically - as we have shown. 
However, in general, the Henry's constants may not describe the single component adsorption behavior well, the adsorbed gas species may interact invalidating the base assumptions of IAST, or there may be more than 2 gas species to consider. 
In this section we show how the Intrinsic DAC cycle analysis could be extended to become a generic Intrinsic Refresh Cycle for these more complicated gas separation problems.
To keep the discussion generic we will refer to a multi-component gas mixture with species: $A$, $B$, $C$, ...

\noindent The initial absolute equilibrium uptake of each species are given by:
\begin{equation}
\begin{aligned}
    n_{1,A} &= n_A(T_1, P_{1,A}, P_{1,B}, P_{1,C}, ...)\\
    n_{1,B} &= n_B(T_1, P_{1,A}, P_{1,B}, P_{1,C}, ...) \\
    n_{1,C} &= n_C(T_1, P_{1,A}, P_{1,B}, P_{1,C}, ...) \\
    ...
\end{aligned}
\end{equation}
\noindent That is, the equilibrium uptake of each species is a function of the temperature and each independent partial pressure. 
While we control the total pressure $P$, the partial pressures are determined by equilibrium with the sorbent. 
Re-writing in terms of $P$ this becomes:
\begin{equation}
\begin{aligned}
    n_{1,A} &= n_A(T_1, x_{1,A}P, x_{1,B}P, x_{1,C}P, ...)\\
    n_{1,B} &= n_B(T_1, x_{1,A}P, x_{1,B}P, x_{1,C}P, ...) \\
    n_{1,C} &= n_C(T_1, x_{1,A}P, x_{1,B}P, x_{1,C}P, ...) \\
    ...
\end{aligned}
\end{equation}
\noindent where:
\begin{equation} \label{eq:genericconstraint}
\begin{aligned}
    x_{n,A} + x_{n,B} + x_{n,C}+ ... = 1 | \forall n
\end{aligned}
\end{equation}
\noindent At the beginning of the intrinsic refresh cycle the system will be in equilibrium with the known composition of the in-coming gas mixture, so the gas compositions $(x_{1,A}, x_{1,B}, x_{1,C}, ...)$ will be known and the equilibrium uptake $(n_{1,A}, n_{1,B}, n_{1,C}, ...)$ can be calculated. 
As the desorption step of the intrinsic refresh cycle begins, at each point along the path through thermodynamic parameters the system will be in equilibrium with the gas that desorbs from the sorbent at the infinitesimally previous point along the path. 
This can be solved for numerically by considering the change in equilibrium uptake. 

\noindent The change in the equilibrium uptakes are:
\begin{equation}
\begin{aligned}
    d_{A} &= n_{2, A} - n_{1, A} \\
    d_{B} &= n_{2, B} - n_{1, B} \\ 
    d_{C} &= n_{2, C} - n_{1, C} \\
    ...
\end{aligned}
\end{equation}

\noindent If there is a known finite volume of the desorption chamber to consider, then the initial moles of each gas species in that finite volume is given by:
\begin{equation}
\begin{aligned}
    m_{1,A} &= (x_{1,A})\frac{PV}{RT} \\
    m_{1,B} &= (x_{1,B})\frac{PV}{RT} \\
    m_{1,C} &= (x_{1,C})\frac{PV}{RT} \\
    ...
\end{aligned}
\end{equation}

\noindent We can then determine the new concentrations with:
\begin{equation} 
\begin{aligned}
    x_{2, A} &= \frac{m_{1, A} + d_{A}}{m_{1, A} + m_{1, B} + m_{1, C} + ... + d_{A} + d_{B} + d_{C} + ...}\\
    x_{2, B} &= \frac{m_{1, B} + d_{B}}{m_{1, A} + m_{1, B} + m_{1, C} + ... + d_{A} + d_{B} + d_{C} + ...}\\
    x_{2, C} &= \frac{m_{1, C} + d_{C}}{m_{1, A} + m_{1, B} + m_{1, C} + ... + d_{A} + d_{B} + d_{C} + ...}\\
    ...
\end{aligned}
\end{equation}

\noindent This can then be re-arranged to give: 
\begin{equation} \label{eq:genericroot}
\begin{aligned}
    0 &= x_{2, A} - \frac{m_{1, A} + d_{A}}{m_{1, A} + m_{1, B} + m_{1, C} + ... + d_{A} + d_{B} + d_{C} + ...}\\
    0 &= x_{2, B} - \frac{m_{1, B} + d_{B}}{m_{1, A} + m_{1, B} + m_{1, C} + ... + d_{A} + d_{B} + d_{C} + ...}\\
    0 &= x_{2, C} - \frac{m_{1, C} + d_{C}}{m_{1, A} + m_{1, B} + m_{1, C} + ... + d_{A} + d_{B} + d_{C} + ...}\\
    ...
\end{aligned}
\end{equation}

\noindent Equation \ref{eq:genericroot} can then be solved with a root finding operation to find the new concentrations under the constraint of Equation \ref{eq:genericconstraint} that the concentrations must sum to 1. 
This processes can then be followed iteratively at finite steps along the refresh path to obtain the equilibrium uptake for the intrinsic refresh cycle. 

\newpage
\section{Refresh Path Calculation Completion Mechanism} \label{sec:CompletionMech}

For each of the 11 660 MOF materials in the CSD database, we applied our Intrinsic DAC cycle analysis using the models described in the main text, and search for Pareto optimal refresh paths. 
As described in the main text, we constrain these paths to be linear and monotonic between two points in $(T,P)$-space. 
We performed GCMC calculations to determine the Henry's constants ($K_H$), then use thermodynamic extrapolation of those to get the temperature dependence. 
We truncate the paths to regions with monotonically decreasing $K_H$ with increasing temperature. 
Since the extrapolated Henery's constants of \ce{CO2} at low temperatures can be very high, we further truncate the paths to where the equilibrium uptake of \ce{CO2} from the Henry's constant isotherm is less than the saturation uptake of \ce{CO2}: $K_{H,CO_2} * P_{CO_2} < n_{CO_2, sat.}$.
We also eliminate non-adsorbing materials, \textit{i.e.} where the \ce{CO2} saturation uptake is zero.
Since the concentration of \ce{CO2} in the approximation of atmospheric conditions is low, and the majority of the total pressure is composed of \ce{N2}, we check that the linear Henry's constant isotherm for \ce{N2} is a good approximation of the equilibrium uptake of \ce{N2}.
We do this by comparing the equilibrium uptake at  one atmosphere of \ce{N2} as predicted by the Henry's constant to a direct GCMC calculation of the \ce{N2} uptake. 
If $n_{N_2} \pm 2* \sigma_{n,N_2}$ from the direct GCMC calculation is within 10 \% of $K_{H,N_2} * P$, then we consider the Henry's constant a good approximation. 
Lastly, we eliminate materials where the Intrinsic DAC cycle optimizer failed.
This can happen during the uncertainty propagation for the equilibrium uptake.
We sample whole trends of $K_H(T)$ from the thermodynamic extrapolation in order to preserve the smoothness and monotonicity of the trend.
Despite this, the sampled trend of $K_H(T)$ can lead to imaginary or complex valued compositions (solutions to Equation \ref{eq:root}) due to numerical issues related to the step size in $T$ affecting the step in $K_H$.

\begin{table}[htp]
    \resizebox{\linewidth}{!}{%
    \begin{tabular}{c|c|c}
    Completion Mechanism & Count  & Likely Cause \\
    \hline \hline  
     \makecell{Materials that failed \\ $K_H$ extrapolation}   & 4 & Insufficient statistics after GCMC trials \\
     \hline
     \makecell{Materials with paths \\ truncated to nothing}    & 2 & \makecell{\{monotonic $K_H(T)$ \} \\ $\cap$ \\ \{ $K_{H,CO_2} * P_{CO_2} < n_{CO_2, sat.}$ \} \\ eliminated all path steps}\\
     \hline
     Non-adsorbing Materials   & 5 & \makecell{Saturation calc. did not adsorb, \\ likely non-porous} \\
     \hline
     \makecell{Materials that failed \\ close-enough test}   & 2880 & \makecell{Non-linear $n_{N_2}(P)$ \\ at 101325 Pa \ce{N2}} \\
     \hline
     \makecell{Materials that failed \\ at optimizer}   & 10 & \makecell{Samples of $K_H(T)$ \\ change too fast for step size \\ leading to imaginary roots \\ of analytical IAST}\\
     \hline
     Successfully optimized materials   & 8759 & \\
     \hline
     Total   & 11660 & \\
     
    \end{tabular}}
    \caption{Table of the completion mechanisms for the Pareto optimization of the refresh paths for the MOFs in the CSD database and the likely cause.}
    \label{tab:CompletionMechanism}
\end{table}